# Six-Pole Dual-band Bandpass Filter for WiMAX Applications

Halimat Olamide Yusuf
*Dept. of Electrical and Electronic*
*School of Engineering*
*University of Greenwich*
London, UK
hy6913o@greenwich.ac.uk

Augustine O. Nwajana
*Dept. of Electrical and Electronic*
*School of Engineering*
*University of Greenwich*
London, UK
a.o.nwajana@greenwich.ac.uk

***Abstract*—Recent advances in multi-band wireless communication systems have driven the increasing need for dual-band bandpass filters. These types of filters are capable of isolating a small section of the frequency spectrum within a broader spectrum. Over the years, coplanar waveguide, microstrip, slotline, stripline, and other planar transmission line technologies have been widely employed in the design of microwave circuits and systems. This work employs a Folded-Arms Square Open-Loop Resonator (FASOLR) microstrip planar structure, designed and simulated using PathWave Advanced Design System (ADS). A third order (three-pole) single bandpass filter is transformed into a sixth order (six-pole) dual-band bandpass filter. The proposed six-pole dual-band bandpass filter is centred at 2.3 GHz, with a fractional bandwidth of 7% and produces two passbands centred at approximately 2.2 GHz and 2.4 GHz. The design is implemented on a commercially available Rogers RT/Duroid 6010LM substrate with a dielectric constant (εr) of 10.7, a loss tangent (tan δ) of 0.0023, a substrate thickness of 1.27 mm, and 35 µm copper cladding on both sides. The overall filter component has a compact footprint of 30.56 mm by 20.56 mm (that is, 0.62 λg × 0.42 λg), with λg indicating the guided wavelength. The design is validated through comparison of circuit-level and electromagnetic (EM) simulation results, which show good agreement. The EM simulation responses indicate a return loss better than 10 dB on the first band and better than 19 dB on the second, and an insertion loss better than 1.5 dB, demonstrating its suitability for WiMAX applications.**

***Keywords—BPF; dual-band; FASOLR; coupling; microstrip; WiMAX***

## I. Introduction

The rapid growth of multi-band wireless communication has placed increasing pressure on the radio-frequency spectrum, driving a sustained demand for compact components capable of operating across multiple bands at once[1], [2]. Bandpass filters play a critical role in addressing this challenge, as they determine which signals are transmitted and which are rejected within a crowded spectral environment[3]. The dual-band bandpass filter (BPF) has emerged as a particularly valuable solution, allowing two distinct frequency bands to be selected within a single component and thereby reducing both the size and the loss of the overall system [4]-[10]. Among planar implementation technologies, microstrip has become the most widely adopted, owing to its compactness, low fabrication cost and compatibility with standard printed-circuit processes [11]-[14].

A wide variety of techniques have been explored to realise dual-band behaviour. Stepped-impedance resonators [11], split-ring resonators [6] and defected microstrip structures [12] generate two passbands through controlled impedance ratios and geometrical perturbation, while dual-mode and coupled-line resonators [5], [14], resonator-stub interaction [2], fractal geometries [13], and degenerate-mode splitting [15] achieve the same end through modal manipulation. Compact realisations based on substrate-integrated-waveguide open-loop rings, coupled-line and stepped-impedance stub networks, and hairpin-ring structures with defected ground planes have similarly advanced miniaturisation and stopband control [16]-[19], often with transmission zeros introduced near the band edges to sharpen selectivity [20]-[22]. Despite this diversity, the reported designs are predominantly realised as lower-order structures, typically second-order per passband, and depend largely on geometrical or modal tuning that offers limited explicit control over transmission-pole allocation. In contrast, the approach adopted in this work converts a three-pole single-band filter into a six-pole dual-band filter, with three transmission poles assigned to each passband. This provides a structured higher-order synthesis route in which the poles of each band are directly derived from the single-band prototype. The same conversion principle can be extended to multi-band filter designs.

In this paper, the proposed technique is demonstrated through a microstrip folded-arms square open-loop resonator (FASOLR) implementation centred at 2.3 GHz for WiMAX applications. The FASOLR is employed owing to its compact geometry and suitability for high-permittivity substrates. It descends from the conventional half-wavelength resonator: bending the half-wavelength line into a square yields the square open-loop resonator (SOLR), whose side measures approximately one-eighth of a guided wavelength, while folding the two arms of the SOLR inwards produces the FASOLR, whose side measures only about one-tenth of a guided wavelength, achieving a further size reduction of roughly 20% relative to the SOLR [23]. This evolution of the resonator structure is shown in Figure 1.

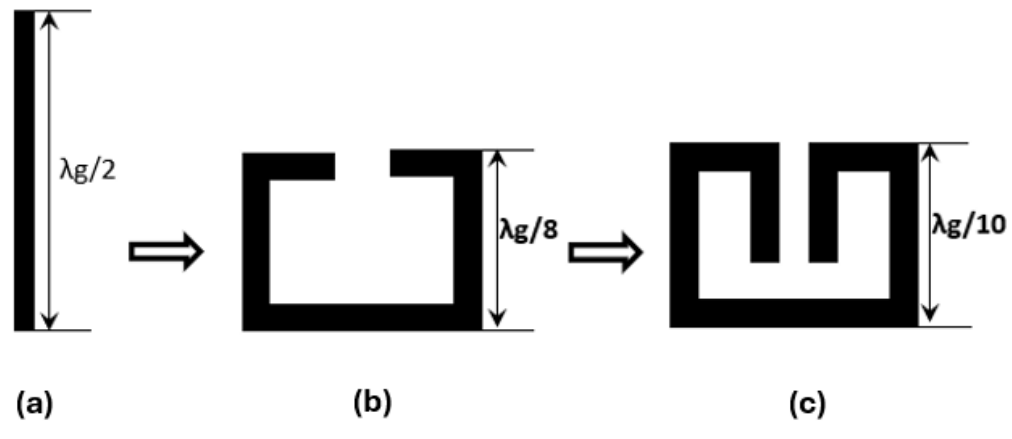

Figure 1. Evolution of the microstrip resonator structure: (a) half-wavelength resonator; (b) square open-loop resonator; (c) folded-arms square open-loop resonator.

## II. THEORETICAL FILTER CONFIGURATION

This section presents the circuit-level synthesis of the proposed filter, in which a three-pole single-band bandpass filter (BPF) is derived from a normalised Chebyshev low-pass prototype and subsequently transformed into the six-pole dual-band configuration. The prototype element values $g_0 = g_4 = 1.0$, $g_1 = g_3 = 0.8516$ and $g_2 = 1.1032$ are transformed into a band-pass network of series and shunt LC resonators, based on the filter design formulations reported in [3], for a centre frequency of 2.3 GHz, a fractional bandwidth of 7%, a passband ripple of 0.04321 dB, and an input/output characteristic impedance $Z_0$ of 50 Ω.

Admittance inverters (J-inverters) are then introduced to realise the coupled-resonator structure and are subsequently replaced by an equivalent capacitor-only network to enable implementation in PathWave Advanced Design System (ADS), following the same formulations [3]. This yields the three-pole single-band capacitor-only configuration shown in Figure 2a, in which the shunt resonators are formed by equal capacitance C and inductance L, and the inter-resonator coupling is provided by the capacitor networks C01 and C12.

The single-band network is then converted into the corresponding six-pole dual-band configuration shown in Figure 2b. The conversion is achieved by extending the circuit, with an additional set of three resonators coupled to the main-line resonators through a similar capacitor-only mutual-coupling network, thereby raising the filter order from three to six and producing the dual-band response. The resulting capacitor and inductor values are C = 16.8368 pF, L = 0.2844 nH, C01 = 1.3840 pF and C12 = 1.2159 pF.

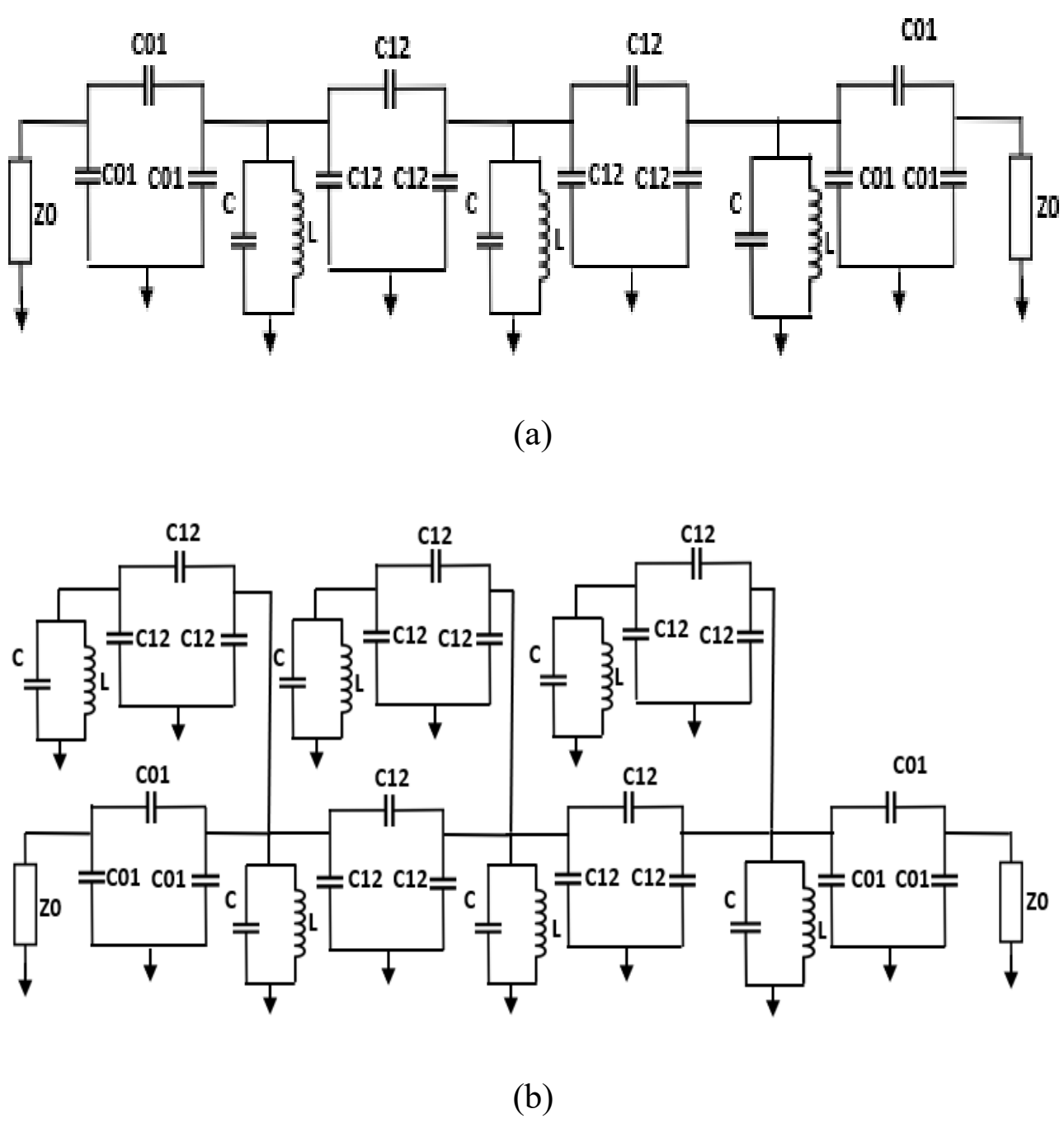

Figure 2. Capacitor-only BPF network: (a) three-pole single-band; (b) six-pole dual-band.

## III. PRACTICAL FILTER CONFIGURATION

The microstrip folded-arms square open-loop resonator (FASOLR) reported in [23] is employed in the implementation of the proposed filter, owing to its compact geometry and suitability for high-permittivity substrates. The resonator width and guided wavelength are determined from [3]. PathWave Advanced Design System (ADS) is used to construct and simulate the layouts, with the coupling arrangements shown in Figure 3. The layouts are implemented on a Rogers RT/Duroid 6010LM substrate with a dielectric constant of 10.7, a thickness of 1.27 mm, a loss tangent of 0.0023, and 35 μm copper cladding on both sides. The FASOLR is designed to resonate at the specified centre frequency of 2.3 GHz.

The proposed six-pole dual-band bandpass filter layout arrangement, shown in Figure 3(b), is derived from the three-pole single-band configuration of Figure 3(a). Three resonators (A, B and C) are first coupled along the main signal path between Port 1 and Port 2 to form the single-band filter. Three additional resonators (A1, B1 and C1) are then coupled to the main-line resonators, raising the filter to six poles and providing the inter-pair coupling that gives rise to the dual-band response. The theoretical coupling coefficients (k) between adjacent resonators (A and B, B and C, A and A1, B and B1, and C and C1) are obtained from the prototype values using Equation (1), while the corresponding external quality factors (Qext) at the input (Port 1 and A) and output (C and Port 2) are determined from Equation (2) [3]. The practical coupling coefficients and Qext are extracted from the EM simulation responses using the frequency-splitting technique reported in [23].

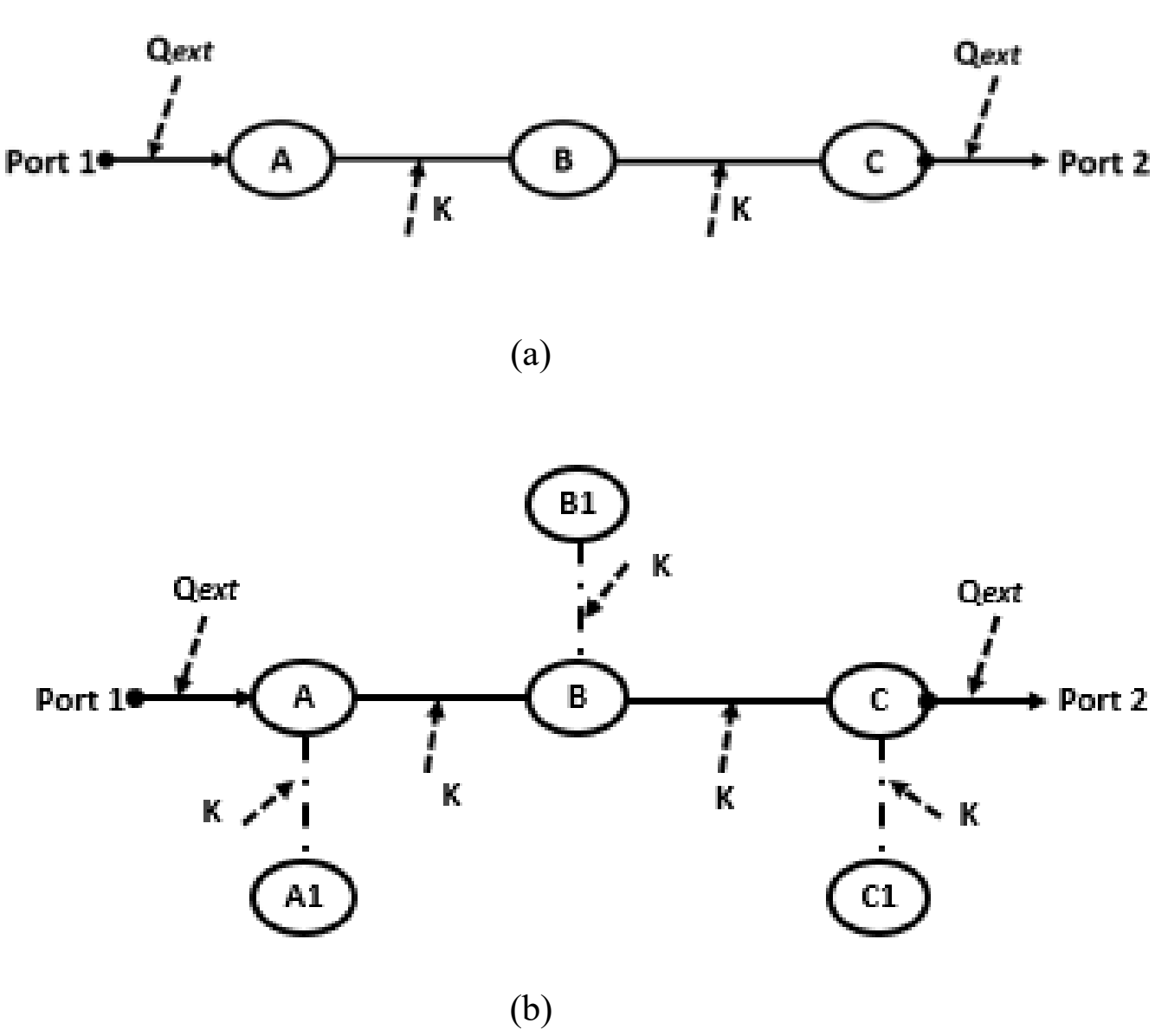

Figure 3. Filter coupling arrangements: (a) three-pole single-band; (b) six-pole dual-band.

$$K = \frac{\mathrm{FBW}}{\sqrt{g_1 g_2}} = 0.0722 \qquad (1)$$

$$Q_{\mathrm{ext}} = \frac{g_0 g_1}{\mathrm{FBW}} = 12.17 \qquad (2)$$

With the coupling arrangements of Figure 3 as the design foundation, the folded-arms square open-loop resonators (FASOLR) are assembled to derive the complete microstrip layout. The physical dimensions of the six-pole dual-band bandpass filter microstrip layout are shown in Figure 4.

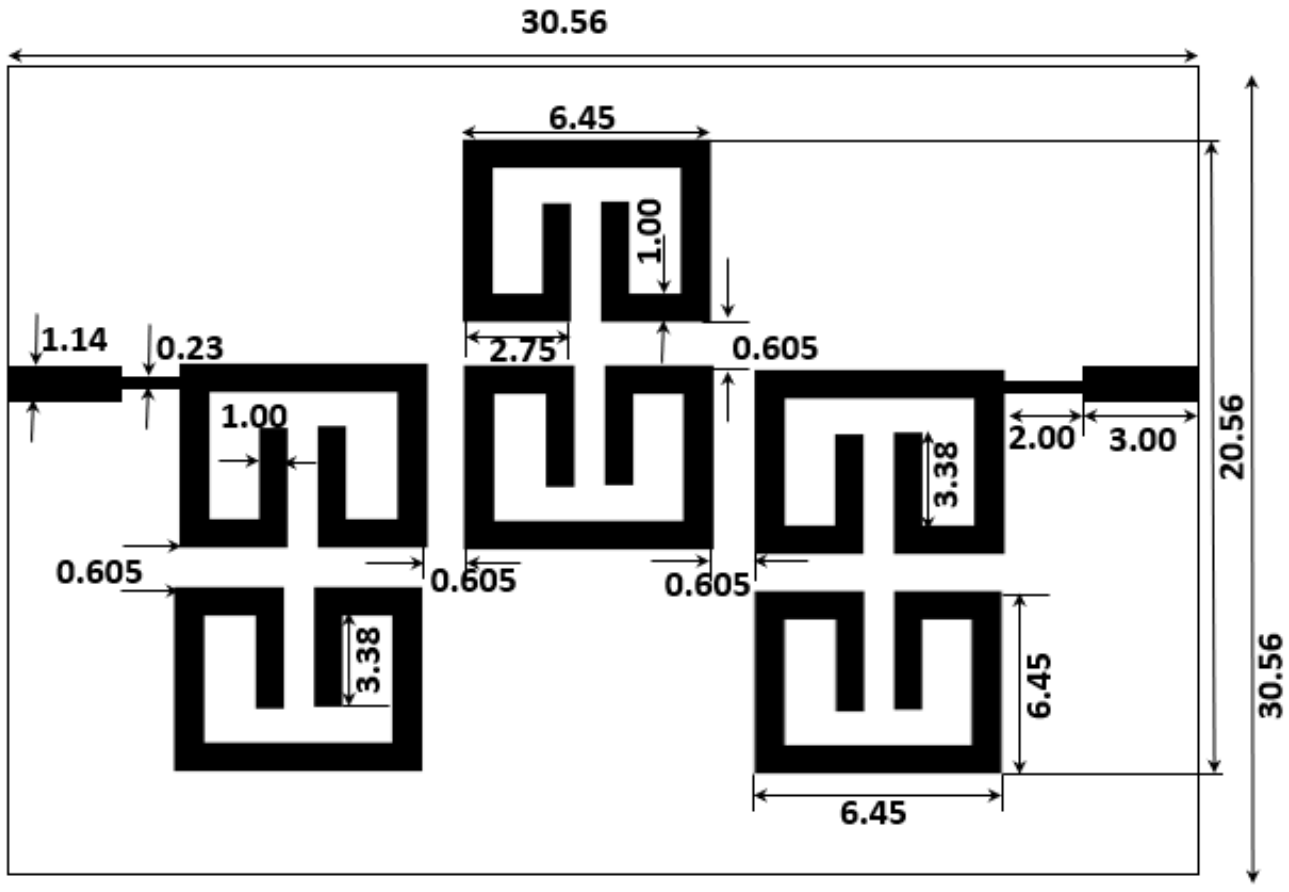

Figure 4. Microstrip layout of the proposed six-pole dual-band bandpass filter (dimensions in mm).

## IV. Results and Discussion

This section presents and discusses the simulation results for the proposed six-pole dual-band bandpass filter. The circuit-level (theoretical) response is examined first, followed by the electromagnetic (EM) (practical) response. A comparison between the theoretical and practical responses is then presented to validate the microstrip realisation.

The theoretical S-parameter response is shown in Figure 5a. The filter exhibits two passbands positioned symmetrically about the 2.3 GHz design frequency, separated by a deep inter-band rejection at the centre. Within each passband, the transmission coefficient ($S_{21}$) approaches 0 dB, while the reflection coefficient ($S_{11}$) presents three minima per band, consistent with the three poles allocated to each band. The low $S_{11}$ levels across both passbands confirm that the capacitor-only coupling network produces the intended dual-band response at the circuit level.

The practical S-parameter response, obtained from full-wave EM simulation of the microstrip layout, is shown in Figure 5b. As in the theoretical case, the filter exhibits two passbands positioned symmetrically about the 2.3 GHz design frequency, separated by a deep inter-band rejection at the centre, with the first passband centred near 2.2 GHz and the second near 2.4 GHz. The return loss remains better than 10 dB across the first passband and better than 19 dB across the second, with an insertion loss better than 1.5 dB in both bands, confirming that the practical microstrip implementation preserves the dual-band response established at the circuit level.

The theoretical and practical responses are compared in Figure 5c and show good agreement, confirming that the microstrip layout reproduces the intended dual-band behaviour. Although the filter was designed for a 7% fractional bandwidth, the realised fractional bandwidths, referenced to the 2.3GHz center frequency, are 5.13% and 6.17% for the theoretical response and 5.09% and 5.48% for the practical response, for the first and second passbands, respectively. The deviation from the 7% design value is attributed to the approximation of the ideal admittance inverters using a capacitor-only network. While the network provides the required inverter behaviour at the design frequency, its frequency-dependent response results in slight variations from the ideal bandwidth specification at the circuit level. For the practical response, the inter-resonator coupling coefficients were obtained from Equation (1) and realised through the physical adjustment of the coupling spacing between resonators in the FASOLR microstrip layout. The realised coupling coefficients were subsequently verified by extraction from the EM simulation using the frequency-splitting technique, showing close agreement with the target values. The residual difference between the realised and target coupling accounts for the deviation from the design bandwidth. The higher insertion loss and shallower rejection of the practical response, relative to the theoretical, are attributed to the conductor and dielectric losses included in the full-wave EM simulation, which are not represented in the idealised, lossless circuit-level model.

The proposed filter is compared with recently reported dual-band designs in Table I. The filter occupies a compact footprint of 0.62 λg × 0.42 λg, where λg is the guided wavelength at the 2.3 GHz centre frequency.

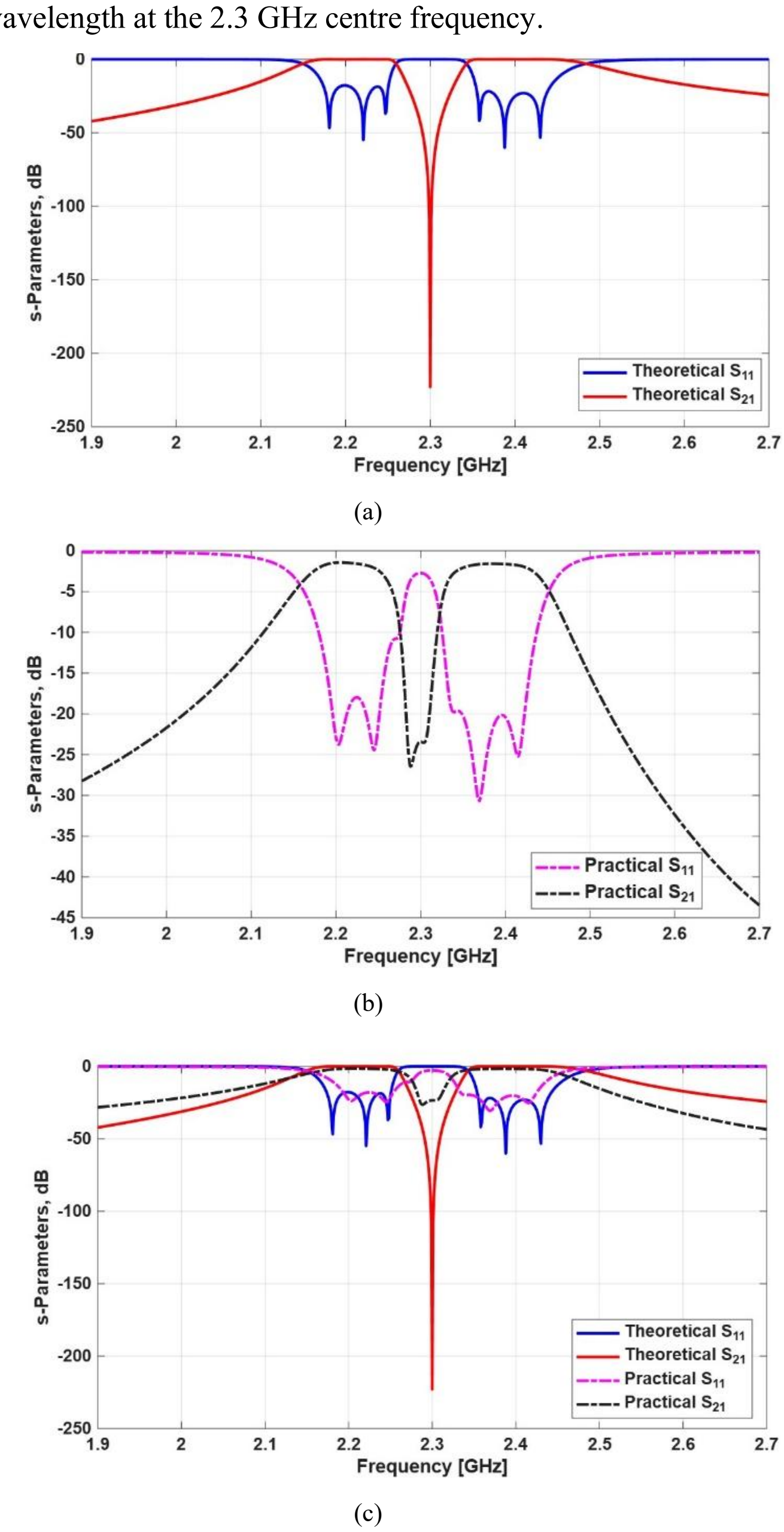

Figure 5. Simulated S-parameter responses of the proposed six-pole dual-band bandpass filter: (a) circuit-level (theoretical) response; (b) electromagnetic (EM) (practical) response; (c) comparison of the theoretical and practical responses.

TABLE I. PERFORMANCE COMPARISON WITH RELATED LITERATURE

| Ref. | $f_1$ / $f_2$ (GHz) | Filter order | Size (λg × λg) | TL[a] | RL[b] (dB) | IL[c] (dB) |
|---|---|---|---|---|---|---|
| [4] | 1.38 / 3.4 | 2 / 2 | 0.33 × 0.20 | MS | 18 / 16.5 | 1.4 / 1.5 |
| [11] | 3.5 / 5.825 | 2 / 2 | 0.527 × 0.639 | MS | 28 | 0.74 |
| [16] | 1.75 / 4.65 | 2 / 2 | 0.041 ($λg^2$) | SIW | 14 / 21 | 1.1/ 1.15 |
| [17] | 0.92 / 1.52 | 3 / 3 | 0.13 × 0.41 | MS | 14 / 16 | 1.01/0.96 |
| This work | 2.2 / 2.4 | 3 / 3 | 0.62 × 0.42 | MS | 10 / 19 | 1.4/1.5 |

[a] transmission line, [b] return loss, [c] insertion loss.

## V. CONCLUSION

A six-pole dual-band bandpass filter for WiMAX applications has been designed and realized using folded-arms square open-loop resonator (FASOLR) technology. The proposed filter is centred at 2.3 GHz and produces two passbands at 2.2 and 2.4 GHz centre frequencies. The design is implemented on a Rogers RT/Duroid 6010LM substrate with a dielectric constant of 10.7, a thickness of 1.27 mm and a loss tangent of 0.0023. The simulated results of the proposed filter show a return loss of better than 10 dB in the first band and better than 19 dB in the second, with an insertion loss better than 1.5 dB across both passbands and a deep rejection notch separating the two bands. The filter occupies a physical size of 0.62 λg by 0.42 λg. From the performance comparison in Table I, the proposed filter offers a higher-order dual-band response, with three poles per band giving sharper selectivity than most reported designs, at a footprint that remains competitive for a six-resonator structure. The proposed filter finds application in modern multifunctional WiMAX communication systems where two closely spaced sub-bands must be selected within a broader spectrum.